\documentstyle[multicol,aps,prl,twocolumn]{revtex}
\newif\iffigs
\figstrue
\iffigs
\fi
\input{psfig}
\def\drawing #1 #2 #3 {
\begin{center}
\setlength{\unitlength}{1mm}
\begin{picture}(#1,#2)(0,0)
\put(0,0){\framebox(#1,#2){#3}}
\end{picture}
\end{center} }

\begin{document}


\twocolumn[\hsize\textwidth\columnwidth\hsize\csname@twocolumnfalse\endcsname

\title{Pdf's of Derivatives and Increments for Decaying Burgers
Turbulence} \author{J. Bec,$^{1}$ and U. Frisch$^1$} \address{$^1$
CNRS UMR 6529, Observatoire de la C\^ote d'Azur, BP 4229, 06304 Nice
Cedex 4, France} \draft
\date{\today} \maketitle
\centerline{{\it Phys. Rev.} E, in press}
\begin{abstract}
A Lagrangian method is used to show that the power-law with a -7/2
exponent in the negative tail of the pdf of the velocity gradient and
of velocity increments, predicted by E, Khanin, Mazel and Sinai (1997
Phys.\ Rev.\ Lett.\ {\bf 78}, 1904) for forced Burgers turbulence, is
also present in the unforced case.  The theory is extended to the
second-order space derivative whose pdf has power-law tails with
exponent -2 at both large positive and negative values and to the time
derivatives. Pdf's of space and time derivatives
have the same (asymptotic) functional forms. This is interpreted in
terms of a ``random Taylor hypothesis''.
\end{abstract}

]

\def\rset{{\rm I\kern -0.2em R}}
\def\un{\hbox{{1\kern -0.25em\raise 0.4ex\hbox{{\scriptsize $|$}}}}}
\def\zset{{\bf Z}}
\def\cset{eq_burg_force\hbox{{C\kern -0.55em\raise 0.5ex\hbox{{\tiny $|$}}}}}
\def\nset{\hbox{{I\kern -0.18em N}}}
\def\la{\left\langle}
\def\ra{\right\rangle}


\section{Introduction}

E, Khanin, Mazel and Sinai \cite{ekhma97} made various predictions
concerning the one-dimensional Burgers equation
\begin{equation}
\partial_t u + u \partial_x u = \nu \partial_{x}^2 u + f,
\label{eq-burg-force}
\end{equation}
with viscosity $\nu$ and a random white-in-time force $f(x,t)$ which
is homogeneous, periodic and smooth in the space variable. One
prediction concerns the probability density function (pdf) of the
velocity gradient $\xi=\partial_x u$. According to
Ref.~\cite{ekhma97}, in the limit $\nu\to 0$, the statistically
stationary solution of (\ref{eq-burg-force}) has a pdf
\begin{equation}
p(\xi) \propto |\xi|^{-7/2}, \quad {\rm for}\,\, \xi \to -\infty.
\label{7halfslaw}
\end{equation}
This power-law range is due to preshocks, nascent shocks with a cubic
root structure, as discussed by Fournier and Frisch
\cite{fofr83}. There has been an interesting controversy about this
negative tail of the pdf, which we shall not try to summarize here
(see, e.g., Refs.~\cite{kra99,eva99a,BFK99} and references therein). There
is no complete proof at this moment of the validity of the -7/2 law,
but significant progress has been made recently
\cite{eva99d}. 
We shall not dwell now on the issue of the validity of the -7/2 law
for  forced {\em burgulence\/} (Burgers turbulence).

It is our intention here to show that the -7/2 law is also present in
unforced decaying burgulence. Specifically, we shall consider
solutions of (\ref{eq-burg-force}) in the limit $\nu\to 0$ with $f=0$
and random zero-mean-value initial conditions $u_0(x)$ which are
periodic (a unit period is assumed for convenience), statistically
homogeneous and sufficiently smooth.  An instance is to take Gaussian
initial conditions with a spectrum decreasing exponentially at high
wavenumbers. Such ``large-scale'' initial conditions will develop
nonsmooth features (preshocks and shocks) after some (random) time.

The paper is organized as follows. In Sec.~\ref{s:lagrangian} we
consider the deterministic problem in Lagrangian coordinates and
identify the preshock events leading to large negative gradients. In
Sec.~\ref{s:fder} we derive the -7/2 law for the pdf of the first
derivative. In Sec.~\ref{s:other}, we derive similar laws for
higher-order space derivatives and the time derivative. In
Sec.~\ref{s:increments} we derive the corresponding results for the
pdf of space increments. In Sec.~\ref{s:conclusion} we make
concluding remarks.

\section{The Lagrangian representation and preshocks}
\label{s:lagrangian}

In the absence of force and of viscous dissipation and as long as no
shock has appeared, the Burgers equation (\ref{eq-burg-force}) has the 
obvious solution
\begin{equation}
u(x,t)=u_0(a), \quad a =L_t^{-1}\,x,
\label{formsol}
\end{equation}
where 
\begin{equation}
L_t: a\mapsto a+tu_0(a),
\label{defnaivelagmap}
\end{equation}
is called the {\it naive Lagrangian map}. This is indeed just a
statement that the velocity of a fluid particle is conserved in
Lagrangian coordinates.(Following standard tradition, we denote
Lagrangian initial coordinates by $a$ and Eulerian
coordinates by $x$.)

A remarkable property of the unforced Burgers equation in the limit
of zero viscosity, which follows from the Hopf--Cole solution (see
Refs.~\cite{sinai,aanda} for details), is that (\ref{formsol}) remains valid in the
presence of shocks provided  the  naive Lagrangian map is replaced 
by the (proper) Lagrangian map ${\cal L}_t$. The latter is defined as
follows. First, we define the  initial potential (up to an additive
constant) by
\begin{equation}
u_0(a) =- \partial_a \psi_0(a).
\label{defpot}
\end{equation}
We then define the  Lagrangian potential
by 
\begin{equation}
\varphi(a,t)\equiv -{a^2\over2}+t\psi_0(a),
\label{deflagpot}
\end{equation}
and observe that the naive Lagrangian map is simply
the negative gradient of the Lagrangian potential\,:
\begin{equation}
L_t\,a=-{\partial\over\partial a} \varphi(a,t).
\label{nlagmaplagpot}
\end{equation}
 The Lagrangian map is defined as
\begin{equation}
{\cal L}_t\,a\equiv -{\partial\over \partial a}\varphi_c(a,t),
\label{deflagmap}
\end{equation}
where $\varphi_c(a,t)$\, is the {\em convex hull} with respect to $a$
of the Lagrangian potential $\varphi(a,t)$.  The convex hull of a
function $f(a)$ can be defined as the smallest piecewise
differentiable function which is greater or equal to $f(a)$ for all
$a$ and such that its derivative is nonincreasing. 

\begin{figure}
\iffigs 
\centerline{\psfig{file=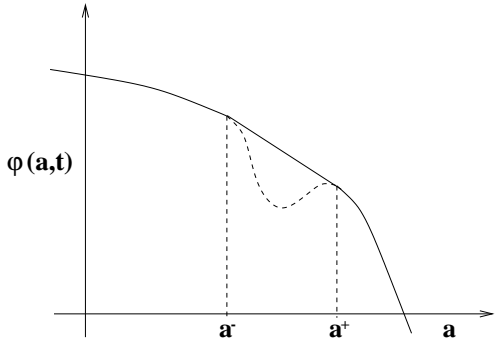,width=6.5cm}}
\else\drawing 65 10 {Lagrangian potential}
\fi
\vspace{2mm}
\caption{Lagrangian potential and its convex hull in the presence 
of a shock interval extending from $a ^-$ to $a ^+$.}
\label{f:convex}
\end{figure}

\begin{figure}
\iffigs 
\centerline{\psfig{file=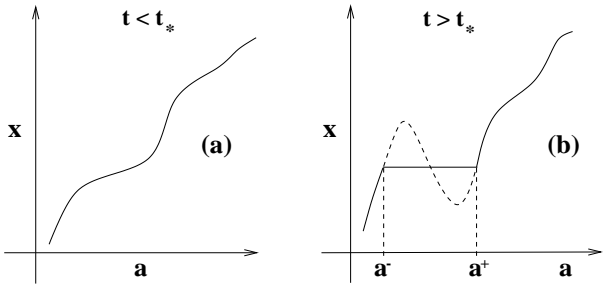,width=8cm}}
\else
\drawing 65 10 {naive and proper Lagrangian maps}
\fi
\vspace{2mm}
\caption{Lagrangian map before (a) and after  (b) the appearance of a
shock. The naive Lagrangian map is shown as a dashed line.}
\label{f:lagmap}
\end{figure}

The graph of the convex hull of $\varphi(a,t)$ is made of pieces of
the graph of the function $\varphi(a,t)$ joined by linear segments,
sitting over the Lagrangian shock intervals, as shown in Fig.~\ref{f:convex}.
Hence, the Lagrangian map coincides with the naive Lagrangian map
except over the Lagrangian shock intervals where it is constant (see
Fig.~\ref{f:lagmap}). Thus, (\ref{formsol}) with $L_t$ given
by (\ref{defnaivelagmap}) remains valid outside the Lagrangian shock intervals.

We can now use this solution (\ref{formsol})-(\ref{defnaivelagmap}) to
calculate the Eulerian velocity gradient, i.e.\ its first-order space
derivative, in Lagrangian coordinates. Differentiating (\ref{formsol})
and using (\ref{defnaivelagmap}), we obtain
\begin{equation}
\partial_x u(x,t) = u'_0(a){1\over \partial_a x}= {u'_0(a)\over 1+tu'_0(a)},
\label{laggrad}
\end{equation}
where $u'_0(a)\equiv du_0(a)/da$. We immediately observe that, for
$t>0$, the only way in which this gradient can become large and
negative is to have a very small denominator in (\ref{laggrad}). 
For the kind of smooth initial conditions considered here, the
denominator is necessarily positive for sufficiently small times. 
Let $a_*$ be the location where  the initial velocity
gradient  $u'_0(a)$ achieves its minimum over the period. At 
\begin{equation}
t = t_* = \min_a\left[ -{1\over u'_0(a)}\right],
\label{deftstar}
\end{equation}
we have the first preshock, i.e. a shock is born
\cite{fofr83}. Subsequently,
other (less negative) local  minima of $u'_0(a)$ may also produce
preshocks, provided  the corresponding location has not already
been captured by a previously generated ``mature shock''. Thus, large negative
Eulerian gradients must come from the neighborhood of preshocks.

We now recall, for use in later sections, the local (normal) form of
the Lagrangian and Eulerian solutions near a preshock. Let $a_*$ be a
local negative minimum of $u'_0(a)$. We then have (generically)
\begin{equation}
u'_0(a_*)<0,\quad u''_0(a_*)=0, \quad u'''_0(a_*)>0.
\label{lestrois}
\end{equation}
Taylor expanding near $a_*$, we
have 
\begin{equation}
u_0(a)\simeq u_0(a_*)+u'_0(a_*)(a-a_*)+{u'''_0(a_*)\over6}(a-a_*)^3.
\label{tayloru}
\end{equation}
By (\ref{defnaivelagmap}),
for $t$ near $t_*= -1/u'_0(a_*)$, 
the naive Lagrangian map is given by
\begin{eqnarray}
x\simeq a_*&+&tu_0(a_*)+{t_*-t\over
t_*}(a-a_*)\nonumber\\
&+&{t_*u'''_0(a_*)\over6}(a-a_*)^3.
\label{taylorx}
\end{eqnarray}
Hence, for given $x$ and $t$, near
$x(a_*,t_*)$ and $t_*$, respectively, the naive Lagrangian map can be
inverted by solving (to leading order) a cubic equation. For $t\le
t_*$, this equation
has a single real solution and the naive Lagrangian map coincides with
the Lagrangian map. For $t=t_*$,
the time of the preshock, the equation simplifies and its solution reads
\begin{equation}
a-a_*\simeq \left[ {6\over t_*u'''_0(a_*)}(x-x(a_*,t_*))\right]^{1/3}.
\label{acubic}
\end{equation}
Substitution in (\ref{tayloru}) gives
\begin{equation}
u(x,t_*)\simeq u_0(a_*) 
-{1\over t_*}\left[ {6\over t_*u'''_0(a_*)}(x-x(a_*,t_*))\right]^{1/3}.
\label{ucubic}
\end{equation}
This is the well known Eulerian cubic root structure of preshocks.
For $t$ slightly in excess of $t_*$, the naive Lagrangian map is not
monotonic and cannot be inverted (otherwise there would be three
branches). There is now a shock. The corresponding shock interval
can be determined by using the convex hull construction on  the 
Lagrangian potential. To leading oder, it is found that the shock
interval extends from $a ^-$ to $a ^+$ which are such that 
$x(a ^{\pm},t)= a_*+tu_0(a_*)$,  namely
\begin{equation}
a ^{\pm}- a_* \simeq\pm \left[{6(t-t_*)\over t_*^2u'''_0(a_*)}\right]^{1/2}.
\label{apm}
\end{equation}
This condition expresses that, to leading order, the Eulerian location
of the shock remains fixed in a frame  moving with the velocity
$u_0(a_*)$. 

\begin{figure}
\iffigs 
\centerline{\psfig{file=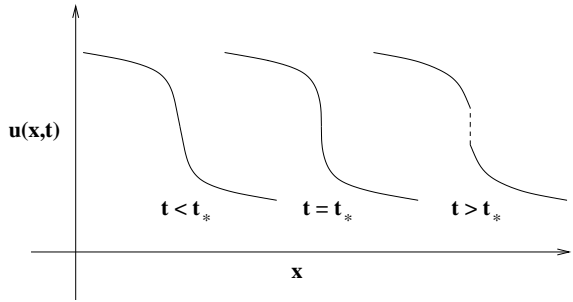,width=7.5cm}}
\else
\drawing 65 10 {preshock, before and after}
\fi
\caption{Eulerian structure of the solution (a) just before a
preshock; (b) at the time of a preshock; (c) just after  a
preshock.}
\label{f:preshock}
\end{figure}

In Fig.~\ref{f:preshock}, we have sketched the Eulerian structure of
the solution at three times, just before, at, and just after the time
$t_*$ of the preshock. 

\section{The PDF of the velocity gradient}
\label{s:fder}

Our purpose is to derive the behavior for $\xi\to -\infty$  of the
pdf of the velocity gradient
\begin{equation}
p(\xi;x,t) \equiv\la \delta(\xi- \partial_xu(x,t))\ra,
\label{defpxi}
\end{equation}
where angular brackets denote ensemble averaging over the random
initial condition $u_0$. By the assumed homogeneity, $p(\xi;x,t)$ is
obviously independent of $x$ (and will subsequently be denoted 
$p(\xi,t)$). It follows  that
\begin{equation}
p(\xi,t)= \la \int_0^1\delta(\xi- \partial_xu(x,t))\,dx\ra.
\label{integral}
\end{equation}
Having thus a representation of our pdf as a space integral over the
Eulerian coordinate $x$, we can make the change of variable from
Eulerian to Lagrangian coordinates, using the map ${\cal L}_t$. The
same idea was used in Ref.~\cite{fofr83} to calculate the Fourier
transform of the Eulerian solution during the early phase of
regularity (before the appearance of shocks).  This idea also works
for later times provided we use the Lagrangian map, which differs from
the naive Lagrangian map (\ref{defnaivelagmap}) only by the exclusion
from the basic periodicity interval $[0,1[$ of the Lagrangian shock
intervals. Let us denote by $D_L(t)$ the set of so-called regular
points, i.e.\ Lagrangian points which do not belong to shock
intervals. Using the Lagrangian representation (\ref{laggrad}) of the
velocity gradient, we obtain from~(\ref{integral})
\begin{equation}
p(\xi,t)= \la \int_{D_L(t)}\delta\left(
\xi -{u'_0(a)\over 1+tu'_0(a)}\right)|1+tu'_0(a)|da \ra.
\label{lagintegral}
\end{equation}
Note that  $1+tu'_0(a)$ is the Jacobian of the Lagrangian map.
Hereafter we shall use several times the formula
\begin{equation}
\delta(f(y))=\sum_j {1\over |f'(y_j)|}\delta(y-y_j),
\label{deltaf}
\end{equation}
where the $y_j$'s are the zeros of $f$ and $f$ is assumed to be
sufficiently smooth. 

Let us denote by $b_k$ the (discrete) Lagrangian locations where
the argument of the delta function in (\ref{lagintegral}) vanishes,
i.e., which are the roots of 
\begin{equation}
\xi- {u'_0(b)\over 1+tu'_0(b)}= 0.
\label{defbj}
\end{equation}
Using (\ref{deltaf}) and evaluating the derivative of the argument of
the delta function in (\ref{lagintegral}) at the point where this
argument vanishes, we can rewrite the pdf of the gradient as
\begin{equation}
p(\xi,t)= {1\over |1-t\xi|^3}\sum_k \la {1\over |u''_0(b_k)|}\int_{D_L(t)}
\delta(a-b_k)\,da \ra,
\label{finexact}
\end{equation}
where the integral over the delta function may be viewed as shorthand
for the indicator function of $D_L(t)$ (equal to one if $b_k\in D_L(t)$
and to zero otherwise). Note that the r.h.s. of (\ref{finexact}) has 
a  $|\xi| ^{-3}$-dependence for large $\xi$, if we take only into
account the first factor. Actually, we shall see that the presence
of $u''_0(b_k)$ in the denominator gives an additional $|\xi|^{1/2}$
factor and that realizability conditions give a $|\xi|^{-1}$ factor,
so that the pdf will be proportional to $|\xi| ^{-7/2}$. 

So far, we have not made any expansion. Let us now concentrate on the
case of large negative $\xi$'s. As observed in
Sec.~\ref{s:lagrangian}, this happens only in the neighborhood of
preshocks. The latter originate from Lagrangian locations at which
$u_0(a)$ has an inflection point satisfying (\ref{lestrois}). Let
$a_{*j}$ be the discrete set of such locations.

We show now by perturbation theory that, for each such point, there
are zero or two roots of (\ref{defbj}). Indeed, using the Taylor
expansion (\ref{tayloru}), in (\ref{defbj}), we obtain
\begin{equation}
(b-a_{*j})^2\simeq {2\over t u'''_0(a_{*j})}\left[-{1\over
t\xi}-1-tu'_0(a_{*j})
\right],
\label{deuxouzero}
\end{equation}
which has either two roots (denoted $b_j^\pm$) or none,
depending on the sign of the r.h.s. Defining $t_{*j}\equiv
-1/u'_0(a_{*j})$, it is now convenient to distinguish the
cases $t\leq t_{*j}$ and $t> t_{*j}$, corresponding respectively to
before and after the preshock. Before, the Lagrangian shock interval
near $a_{*j}$ is empty; the two conditions that $t\leq t_{*j}$ and that 
the r.h.s. of (\ref{deuxouzero}) be positive  read
\begin{equation}
-{1\over t}\leq u'_0(a_{*j}) < -{1\over t} -{1\over t^2\xi}.
\label{avant}
\end{equation}
After $t_{*j}$, the Lagrangian shock interval is defined by
(\ref{apm}). Since shock intervals are excluded from the integral 
(\ref{finexact}), acceptable
solutions must be outside of such  intervals. This and $t>t_{*j}$
gives again two conditions, namely
\begin{equation}
-{1\over t} +{1\over 2t^2\xi}<  u'_0(a_{*j}) < -{1\over t}.
\label{apres}
\end{equation}
We observe now that the two conditions (\ref{avant}) and (\ref{apres})
may be written as a single condition
\begin{equation}
-{1\over t} +{1\over 2t^2\xi}<  u'_0(a_{*j}) < -{1\over t} -{1\over t^2\xi},
\label{avantapres}
\end{equation}
which, for large negative $\xi$, restricts $u'_0(a_{*j})$ to being
near $-1/t$ in a small interval of length  $-3/(2t^2\xi)$.  We shall
denote  by $\un_{PS}(u'_0(a_{*j});t,\xi)$ the indicator
function equal to one if $u'_0(a_{*j})$ is in this ``preshock''
realizability interval and to zero otherwise. 

The integral over $a$ appearing in (\ref{finexact}) is over the
complement of shock intervals. The above analysis takes care only of
those nascent shock intervals which can be calculated
perturbatively. A further condition is that the point $a_{*j}$ should
not be within a mature shock interval which was created before time
$t$. Because of the convex hull construction, this is a global
geometrical constraint which cannot, in general, be expressed
perturbatively. We shall denote by $\un_{D_G(t)}(a)$ the indicator
function equal to one if $a$ is outside such a ``global'' shock
interval and to zero otherwise.

We now return to (\ref{finexact}). Since $b_k=b_j^\pm$, which is close
to the $a_{*j}$'s where $u''_0$ vanishes, we can use the Taylor
expansion $|u''_0(b_j^\pm)|\simeq |b_j^\pm -a_{*j}|u'''_0(a_{*j})$,
which by (\ref{deuxouzero}) takes the same value for $b_j^+$ and
$b_j^-$.  Hence, the contribution to the integral in
(\ref{finexact}) of the two $b_k$ points in the
neighborhood of each point $a_{*j}$  is 
$2\,\un_{PS}\left(u'_0(a_{*j});t,\xi\right)\un_{D_G(t)}(a_{*j})$, where the
products of indicator functions takes care of the shock interval
exclusion.
Using (\ref{deuxouzero}), we now obtain
\begin{eqnarray}
&&p(\xi,t)\simeq {t^{1/2}\over |t\xi|^3} \times \nonumber\\
&&\la \sum_j
{2\,\un_{PS}(u'_0(a_{*j});t,\xi)\un_{D_G(t)}(a_{*j})\over 
\left\{2 u'''_0(a_{*j})\left[-{1\over
t\xi}-1-tu'_0(a_{*j})
\right]\right\}^{1/2}}\ra.
\label{oulabete}
\end{eqnarray}
Since the sum is over points of vanishing $u''_0$ with $u'''_0>0$, by use of
(\ref{deltaf}) this may be rewritten as
\begin{eqnarray}
&&p(\xi,t)\simeq {(2t)^{1/2}\over |t\xi|^3} \Biggl\langle \int_0^1da 
{(u'''_0(a))^{1/2}\delta\left(u''_0(a)\right)
\over \left[-{1\over t\xi}-1-tu'_0(a) \right]^{1/2}}\times \nonumber\\
&& 
\,\,\,\,\times {\rm H}(u'''_0(a))\,\un_{PS}(u'_0(a);t,\xi)
\un_{D_G(t)}(a)
\Biggr\rangle,
\label{oulabebete}
\end{eqnarray}
where ${\rm H}(\cdot)$ is the Heaviside function. 

Interchanging now the mean value and the integration over $a$, we
observe that, because of homogeneity, the integrand does
not depend on $a$. Hence, the integration over $a$ can be omitted.
Let us denote by $p_{3,0}(u',u'',u'''|\un_{D_G(t)}=1)$ the joint pdf of
the first three derivatives of the {\em initial\/} velocity at an arbitrary
Lagrangian location, knowing that this location is not within a mature
shock interval at time $t$. We can then write
\begin{eqnarray}
&&p(\xi,t)\simeq {(2t)^{1/2}\over |t\xi|^3}  \int_0^\infty du'''
\times \nonumber\\
&&\int_{-{1\over t} +{1\over 2t^2\xi}}^{-{1\over t} -{1\over t^2\xi}} du'\,
{(u''')^{1/2}p_{3,0}\left(u',0,u'''|\un_{D_G(t)}=1\right)\over \left[-{1\over t\xi}-1-tu'\right]^{1/2}}.
\label{ouflabebete}
\end{eqnarray}
In (\ref{ouflabebete}), the variable $u'$ is constrained to remain
very close to $-1/t$ for $\xi\to -\infty$. Assuming that the density
$p_{3,0}$ is smooth in its $u'$ argument, we can replace the latter by
$-1/t$ and carry out the remaining integration over $u'$, to obtain
\begin{equation}
p(\xi,t)\simeq 2\sqrt3 t^{-4}D(t) \, |\xi|^{-7/2}, \quad \xi \to -\infty,
\label{lavoici}
\end{equation}
where 
\begin{equation}
 D(t) \equiv \int_0^\infty du'''(u''')^{1/2}p_{3,0}\left(-{1\over
t},0,u'''\,\bigl|\, \un_{D_G(t)}=1\right).
\label{constantD}
\end{equation}
This concludes the derivation of the -7/2 law for decaying burgulence.
The time-dependent constant $D(t)$ is expressed in terms of a
conditional probability which cannot be calculated without solving a
global geometrical random problem \cite{positive}.  A more explicit
form is obtained for small $t$\,: the condition $\un_{D_G(t)}=1$ may
then be omitted and the integral (\ref{constantD}) can be calculated,
e.g., in the Gaussian case. Indeed, for the kind of large-scale
initial conditions assumed here if, near some point $a_*$, the initial
velocity gradient $u'_0$ achieves a very large negative minimum close
to $-1/t$, the other minima will be above $-1/t$ with a probability
very close to unity, so that it is close to certain that no mature
shocks have been formed.

\section{Higher-order space derivatives and time derivatives}
\label{s:other} 

We begin with the second Eulerian space derivative $\partial^2_x
u$. The method is rather similar to the one used for the first
derivative. So, we shall avoid repeating details.  From (\ref{formsol}),
(\ref{defnaivelagmap}) and (\ref{laggrad}), it follows that
\begin{eqnarray}
\partial^2_x u& =& \partial_a (\partial_x u){1\over \partial_a
x}\nonumber\\
& =& \partial_a\left[{u'_0(a)\over 1+tu'_0(a)}\right]{1\over
1+tu'_0(a)}
= {u''_0(a)\over [1+tu'_0(a)]^3}.
\label{derseceul}
\end{eqnarray}
Denoting by $p^{(2)}(\xi,t)$ the pdf of 
$\partial^2_x u$, we have, as before
\begin{equation}
p^{(2)}(\xi,t)= \la \int_{D_L(t)}\!\!\delta\left(
\xi - {u''_0(a)\over [1+tu'_0(a)]^3}\right)|1+tu'_0(a)|da \ra.
\label{lagintegralsecder}
\end{equation}
For Gaussian statistics or, more generally, when the probability of
very large values of $u''_0(a)$ is very small, large
values of $\partial^2_x u$ will be due overwhelmingly to small
denominators.  That is, they will again originate from the
neighborhood of preshocks. Near an inflection point $a_{*j}$ of the
kind considered in Sec.~\ref{s:fder}, we have
\begin{equation}
\partial^2_x u \simeq {u'''_0(a_{*j})(a- a_{*j})\over 
\left[1+tu'_0(a_{*j}) +{t\over2}u'''_0(a_{*j})(a- a_{*j})^2\right]^3}.
\label{dersecpert}
\end{equation}
This is an odd function of $a- a_{*j}$ which can achieve both large
positive and negative values. (Hence, we obtain power-law tails
at both ends.)

As before, in (\ref{lagintegralsecder}) we change from a delta
function over $\xi$ to delta functions over those Lagrangian locations
$b_k$ where the second space derivative is equal to $\xi$. It is
easily shown that the condition that the r.h.s. of (\ref{dersecpert})
be equal to $\xi$ has either two solutions (on the same side of
$a_{*j}$) or none. Obtaining these solutions explicitly requires
solving an algebraic equation of degree six in
$b-a_{*j}$. Nevertheless, the conditions for existence of such
solutions can be written explicitly, as before. In the early shock
phase there may now be either two, or one or zero $b_k$'s outside of
the shock interval. The length of the realizability interval in the
variable $u'_0(a_{*j})$ is now $O(|\xi|^{-2/5})$. As to the
coefficients in front of the distributions $\delta(a-b_k)$, it is now
found to be $O(|\xi|^{-8/5})$. It follows that
\begin{equation}
p^{(2)}(\xi,t) \propto |\xi|^{-2}, \quad \xi \to \pm \infty.
\label{resulp2}
\end{equation}
The time-dependent constant in front the -2 power law can again be
expressed in terms of the conditional joint probability $p_{3,0}$
already introduced, but this is very cumbersome since it involves the
solution of the aforementioned equation of degree six.

The theory can be extended to higher-order space derivatives but
becomes even more cumbersome. Somewhat superficial inspection (mostly by
dimensional analysis) indicates that the pdf's have then  power-law tails
with exponent $-(3n+4)/(3n-1)$ \cite{khaninpc}. For even $n$ the tail
is present for both large negative and positive values. For odd $n>2$
it is certainly present for large negative values and may also be
present for positive ones (e.g. for $n=3$).

Finally, we turn to the Eulerian time derivative. We define
\begin{equation}
p_{\partial_tu}(\eta,t)= \la \delta\left(
\eta - \partial_t u(x,t)\right)\ra.
\label{defpeta}
\end{equation}
From (\ref{formsol}) we
have
\begin{equation}
\partial_tu(x,t)= u'_0(a)\partial_t a,
\label{dudt}
\end{equation}
where $\partial_t a$ is calculated for a given Eulerian position $x$.
Time differentiation of $x=a+tu_0(a)$ gives
\begin{equation}
\partial_t a = -{u_0(a)\over 1+tu'_0(a)}.
\label{dat}
\end{equation}
Hence, 
\begin{equation}
\partial_tu(x,t)= -{u_0(a)u'_0(a)\over 1+tu'_0(a)}.
\label{dtuuprime}
\end{equation}
Note that the r.h.s. is just $-u\partial_x u$, as could have been 
deduced from the Eulerian inviscid equation.
Substituting (\ref{dtuuprime}) in (\ref{defpeta}) and proceeding
almost exactly as in Sec.~\ref{s:fder}, we obtain
\begin{equation}
p_{\partial_t u}(\eta,t)\simeq 2\sqrt3 t^{-4} E^{\pm}(t) \, |\eta|^{-7/2},
\quad \eta \to \pm\infty,
\label{lavoicidt}
\end{equation}
where
\begin{eqnarray}
E^{\pm} (t) &\equiv \pm \int_0^{\pm\infty}du |u|^{5/2}\int_0^\infty 
du'''(u''')^{1/2}\times \nonumber \\
&\times p_{4,0}\left(u,-{1\over
t},0,u'''\,\bigl|\, \un_{D_G(t)}=1\right),
\label{constantDdt}
\end{eqnarray}
which involves the joint pdf $p_{4,0}(u,u',u'',u'''|\un_{D_G(t)}=1)$ of
the initial velocity and its first three derivatives at an arbitrary
location, knowing that this location is not within a mature shock
interval at time $t$. Note that the pdf of the time derivative
(\ref{lavoicidt}) is just the pdf of the space
derivative(\ref{lavoici}) with the change of variable $\eta \to
-u_0\xi$ and an extra averaging over $u_0$. This is the result we
expect if, in (\ref{dtuuprime}), we neglect the variation of $u_0(a)$
near a preshock. It is easily shown, when doing the complete
asymptotic expansion along the same lines as in Sec.~\ref{s:fder},
that this is indeed the case for the leading-order behavior.
This theory can again be extended to pdf's of
higher-order time derivatives which follow the same power laws
as for space derivatives.

Obtaining for the pdf of the Eulerian time derivative the same law as
for the space derivative is not very surprising. In high-Reynolds
number turbulent flows it is well known that, when there is a large
mean flow, the Eulerian temporal structure is, to leading order,
determined by the spatial structure in the reference frame of the mean
flow (this is often referred to as the ``Taylor hypothesis'', but is
of course a simple asymptotic result). Furthermore, when there is no
mean flow, it is generally believed that the small-scale temporal
structure is still determined by the spatial structure, since most of
the time dependence comes from the sweeping of small-scale eddies by
larger energy-containing eddies which have much larger but random
velocities. For the case of burgulence the identical functional forms
of (\ref{lavoici}) and (\ref{lavoicidt}) may be seen as a proof of
this ``random Taylor hypothesis''. Note that it is the sweeping by the
random velocities of the shocks [$u_0(a)$ at those locations where
$u'_0(a)<0$, $u''_0(a)=0$ and $u'''_0(a)>0$] which determines the
interplay of temporal and spatial structures. Since we assumed that
the velocity has zero mean value, the random velocities at the shocks
can have both signs, so that the -7/2 power law tail appears both at
large positive and large negative values of the Eulerian time
derivatives.  Alternatively, one may calculate the pdf of $\partial_t
u$ in the frame moving with the shock (assuming there is a single
shock). In this case one obtains a much steeper law $\propto
|\eta|^{-6}$. Note that this is not the pdf of the Lagrangian time
derivative. For unforced burgulence in the inviscid limit, this
derivative is exactly zero.

\section {Velocity increments}
\label{s:increments}

We define the Eulerian velocity increment over a separation $\Delta x$
as
\begin{equation}
\Delta u_E(\Delta x;x,t) \equiv u(x+\Delta x,t) -u(x,t).
\label{defdeltaueul}
\end{equation}
Our goal is to find the pdf 
\begin{equation}
p_{\Delta u}(\xi,\Delta x,t) \equiv \la \delta\left(\xi - 
\Delta u_E(\Delta x;x,t)\right)\ra,
\label{defpdeltau}
\end{equation}
for values $\xi$ of sign opposite to that of the separation. Let us
introduce the Lagrangian velocity increment, defined outside the
Lagrangian shock intervals, as
\begin{equation}
\Delta u_L(\Delta x;a,t)\equiv \Delta u_E(\Delta x;a+tu_0(a),t).
\label{defdeltaulag}
\end{equation}
Proceeding as at the beginning of Sec.~\ref{s:fder}, we obtain
\begin{eqnarray}
\lefteqn{ p_{\Delta u}(\xi,\Delta x,t) =} \nonumber \\
&&\la \int_{D_L(t)}\!\!\delta\left(\xi-
\Delta u_L(\Delta x;a,t)\right)  
 |1+tu'_0(a)|da \ra.
\label{lagintpdeltau}
\end{eqnarray}
For given $t$, $\Delta x$ and $\xi$, we must now find those Lagrangian
locations, denoted $b_k$, where the argument of the delta function 
in (\ref{lagintpdeltau}) vanishes. For this, it is convenient to
associate to each  $b_k$, the point $b'_k$ such that their images by
the Lagrangian map ${\cal L}_t$ are separated by a distance $\Delta
x$, while the velocities differ by $\xi$.
We thus have
\begin{eqnarray}
u_0(b'_k)-u_0(b_k)&=&\xi \label{deltauxi}\\
b'_k+tu_0(b'_k) &=& b_k+tu_0(b_k)+\Delta x.
\label{bmoinsbdeltax}
\end{eqnarray}
The equivalent of (\ref{finexact}) is now
\begin{eqnarray}
\lefteqn{ p_{\Delta u}(\xi,\Delta x,t) =} \nonumber \\
&&\sum_k\la \left|{(1+tu'_0(b'_k))(1+tu'_0(b_k))\over 
u'_0(b'_k)-u'_0(b_k)}\right|\int_{D_L(t)}\delta(a-b_k)da \ra.
\label{finexactdeltau}
\end{eqnarray}

We shall here be interested exclusively in situations where
\begin{equation}
|\Delta x| \ll |t\xi|\ll 1,
\label{universal}
\end{equation}
which originate from the neighborhood of preshocks $a_{*j}$ where the
Taylor expansion 
(\ref{tayloru}) may be used \cite{remshraiman}. From (\ref{deltauxi}) and
(\ref{bmoinsbdeltax}), we then obtain that $b_k$ is a root of the
following quadratic equation in $b$\,:
\begin{eqnarray}
&&(b-a_{*j})^2 +t\xi(b-a_{*j})+{t^2\xi ^2\over 3}\nonumber\\
&&+{2\over tu'''_0(a_{*j})}\left[1+tu'_0(a_{*j})+{\Delta x\over t\xi}\right]=0.
\label{quadratic}
\end{eqnarray}
We shall see that (\ref{quadratic}), together with the realizability
condition of having real roots not belonging to a shock interval, can
have either zero, one or two solutions, which we shall denote by
$b_{jm}$.  We then approximate the pdf (\ref{finexactdeltau}) using
the Taylor expansion (\ref{tayloru}) near preshocks, to obtain
\begin{eqnarray}
\lefteqn{ p_{\Delta u}(\xi,\Delta x,t) \simeq} \nonumber \\
&&\sum_{jm}\la \left| A_{jm}+B_{jm}\right|\int_{D_L(t)}\delta(a-b_{jm})da \ra,
\label{pdevelope}
\end{eqnarray}
where
\begin{eqnarray}
A_{jm} &=&
{\left[1+tu'_0(a_{*j})+tu'''_0(a_{*j})(b_{jm}-a_{*j})^2/2\right]^2
\over t\xi u'''_0(a_{*j})(b_{jm}-a_{*j})},\\
B_{jm} &=& t\left[1+tu'_0(a_{*j})+tu'''_0(a_{*j})(b_{jm}-a_{*j})^2/2\right].
\label{horreur}
\end{eqnarray}

The realizability conditions associated to (\ref{quadratic}) lead to 
rather involved expressions for the pdf. Simple scaling behavior
emerges in two limiting cases. To express the corresponding
conditions in reasonably compact fashion,
we shall assume that the third derivative $u'''_0(a_{*j})$ is of order
unity. 

For $|t\xi|\ll |\Delta x|^{1/3}$ the $A_{jm}$ term in (\ref{pdevelope})
dominates. For $|t\xi|\gg |\Delta x|^{1/3}$, the  $B_{jm}$ term
dominates. In the former case, $A_{jm}$ can be further approximated
\begin{equation}
|A_{jm}|\simeq {|\Delta x|^2\over |t\xi|^3|u'''_0(a_{*j})(b_{jm}-a_{*j})|}.
\label{dejavu}
\end{equation}
(Realizability imposes that $\xi$ and $\Delta x$ be of opposite sign.)
Substitution of (\ref{dejavu}) into (\ref{pdevelope}) leads
essentially
to the same expression (\ref{oulabete}) as for the pdf of velocity
gradients, provided (i) we replace the Eulerian velocity gradient
by $\xi/\Delta x$ and (ii) multiply the pdf (\ref{oulabete}) by $1/|\Delta
x|$.
Hence,
\begin{eqnarray}
&&p_{\Delta u}(\xi,\Delta x,t) \simeq {1\over |\Delta
x|}\,\,p\left({\xi\over \Delta x},t\right)
 \propto  |\Delta x|^{5/2}|\xi|
^{-7/2},\label{Ado1}\\&& {\rm for}\,\,|\Delta x|\ll |t\xi|\ll |\Delta
x|^{1/3},\,\,\,{\xi\over\Delta x} <0.
\label{Ado2}
\end{eqnarray}

For $1\gg |t\xi|\gg |\Delta x|^{1/3}$ the $B_{jm}$ term in
(\ref{pdevelope}) dominates. Contrary to the former case, the
situation is quite different from what has been studied in
Sec.~\ref{s:fder}. We shall thus give more detail. With the assumptions
made, the condition that the quadratic equation (\ref{quadratic}) have
real roots reads
\begin{equation}
u'_0(a_{*j}) \leq -{1\over t}-{u'''_0(a_{*j})\over 24} t^2 \xi ^2,
\label{discriminant}
\end{equation}
which, given the positivity of $u'''_0(a_{*j})$, implies that
$t>t_{*j}= -1/u'_0(a_{*j})$. It is then easily checked that one 
of the two roots of (\ref{quadratic}) is not acceptable because it
is within the Lagrangian shock interval. The condition that the
other one be outside this interval reads
\begin{equation}
 -{1\over t}-{u'''_0(a_{*j})\over 6} t^2 \xi ^2\leq u'_0(a_{*j}).
\label{leftoutside}
\end{equation}
Eqs.~(\ref{discriminant}) and (\ref{leftoutside}) play now the role of
the realizability conditions (\ref{avantapres}) in Sec.~\ref{s:fder}.
Proceeding then along the same lines as in Sec.~\ref{s:fder}, we
obtain
\begin{eqnarray}
&&p_{\Delta u}(\xi,\Delta x,t) \simeq {t^3\over 8} F(t)|\Delta x|\,|\xi|,
\label{Bdo1}\\
&& {\rm for}\,\,|\Delta x|^{1/3}\ll |t\xi|\ll 1,\,\,\,{\xi\over\Delta x} <0,
\label{Bdo2}
\end{eqnarray}
where
\begin{equation}
 F(t) \equiv \int_0^\infty du'''(u''')^2\,p_{3,0}\left(-{1\over
t},0,u'''\,\bigl|\, \un_{D_G(t)}=1\right),
\label{constantF}
\end{equation}
and $p_{3,0}$ is defined as in  Sec.~\ref{s:fder}.

We observe that the $\xi$-dependence of our results (\ref{Ado1}) and
(\ref{Bdo1}) for decaying burgulence is essentially the same as what
was proposed in Ref.~\cite{ekhma97} for the forced case.

\section{Concluding remarks}
\label{s:conclusion}

We have shown here that several results proposed in Ref.~\cite{ekhma97}
for forced burgulence are also valid for decaying burgulence and can
actually be derived by systematic asymptotic expansions, using a
Lagrangian approach. The results which carry over from the forced to
the unforced case are those involving preshocks\,: -7/2 power law for
the pdf of velocity gradients and increments and +1 power law for the
pdf of increments over suitable ranges.  For the tail of the pdf of
gradients at large {\em positive\/} $\xi$ a decaying exponential law of the
argument $\xi ^3$ is generally proposed
\cite{ekhma97,pol95,gumi96,bafako97,gokr98}.  This result, unrelated
to preshocks, has no counterpart in the decaying case.  Indeed, it
follows then from $D\xi/Dt=-\xi ^2$, where $D/Dt$ denotes the
Lagrangian derivative, that the pdf of $\xi$ is exactly zero for
$\xi>1/t$.

Our results about the -7/2 power law are quite explicit\,: for
example we obtain the (time-dependent) constant $D(t)$, given by
(\ref{constantD}), in front of the power law (\ref{lavoici}). The
expression of $D(t)$ for short times, when mature shocks have
negligible probability, can be written explicitly in terms of the
joint pdf of the first three derivatives of the initial velocity 
at an arbitrary location. For later times we need the
conditional pdf knowing that no mature shock is present at that
location. Obtaining this exactly, say for Gaussian
initial conditions, may be very hard. But it is possible to construct
lower bounds. For example, large deviations theory may be used to
show that  $-\ln D(t)= O(\ln ^2t)$ for $t\to \infty$ \cite{molchan}.

We observe that, formally, our results can easily be extended from the
case of a space-periodic homogeneous initial condition $u_0(a)$ (as
assumed in Ref.~\cite{ekhma97}) to that of a random homogeneous mixing
initial condition defined on the whole real line. For this it suffices
to use ergodicity and to replace, in (\ref{integral}), the mean of the
integral over the period by
$\lim_{L\to\infty}(1/(2L))\int_{-L}^{+L}$. After this, the calculation
is essentially unchanged.

It is of interest to point out that without careful handling of the
shock conditions, an incorrect $|\xi|^{-3}$ power law is obtained for
the left tail of the pdf of gradients. Indeed, the pdf
(\ref{finexact}) has an overall $|\xi|^{-3}$ factor in front of the
r.h.s. (for large $|\xi|$). If in the remaining factor we perform the
integration over the whole Lagrangian interval $[0,1[$ without
excluding the Lagrangian shock interval, we obtain an order unity
contribution. It has already been pointed out in Ref.~\cite{eva99b}
that a $|\xi|^{-3}$ law is obtained from multi-valued solutions
of the Riemann equation. This is indeed equivalent to using the
naive Lagrangian map; note that it gives the correct answer when using
the Zeldovich approximation in cosmology, which allows
``multi-stream'' solutions \cite{Zeld,FBV99}.

It is shown in Ref.~\cite{libro} that there is a simple relation
between the velocity gradient and the density of an advected passive
scalar. When this density is initially uniform, this implies that the
power law with exponent $-7/2$ also applies to the tail of the density
pdf. Analogous results can be obtained in several dimensions, where
they have cosmological implications; this requires the study of
singularities of multi-dimensional convex hulls \cite{FBV99}.

Finally, there is a problem which is in a way intermediate between
the decaying and the forced case, namely ``kicked'' burgulence. The
space-periodic force $f$ appearing in (\ref{eq-burg-force}) is then of the form
\begin{equation}
f(x,t) = \sum_j f_j(x)\delta(t-t_j),
\label{defkicked}
\end{equation}
where the $f_j(x)$ are deterministic or random prescribed functions.
Between the ``kicking times'' $t_j$ we have decaying dynamics. At
time $t_j$ the velocity  undergoes a temporally (but not spatially)
discontinuous change $f_j(x)$. When the $t_j$'s are equally spaced
and all the $f_j(x)$'s are equal, the solution of the Burgers equation
converges to a space-time-periodic function. Pdf's obtained by 
space and time averaging have  exactly the same scaling properties as 
obtained here for random decaying burgulence. Very clean scaling can
be obtained by numerically simulating this problem using a
modification of the fast
Legendre transform method of Refs.~\cite{aanda,nouve94}. Such questions
are discussed in Ref.~\cite{BFK99}.

We thank M.~Blank, S.~Chen, W.~E, S.~Gurbatov, K.~Khanin,
R.~Kraichnan, A.~Noullez, B.~Shraiman, E.~Vanden Eijnden and B.~Villone for
fruitful discussions.



\begin{references}
\bibitem{remshraiman} We are thereby limiting ourselves to the
consideration of small-amplitude shocks which are newborn. Small shock
amplitudes can also be obtained when a shock has decayed for a time
long compared to its time of creation; for times order unity, the
latter phenomenon happens only with a very small probability and does
not contribute to the leading-order expression of the pdf of velocity
increments over small separations.
\end{references}
\end{document}